\documentclass{PoS}
\usepackage{cite}
\usepackage{ textcomp }
\usepackage{ amssymb }
\usepackage{ mathrsfs }
\usepackage{lineno}

\title{Top quark pair property measurements using the ATLAS detector at the LHC}

\ShortTitle{Top quark pair property measurements using the ATLAS detector at the LHC}

\author{
	\speaker{Mohammad J. Kareem}%
         \thanks{On behalf of the ATLAS Collaboration}\\
       II. Physikalisches Institut, Georg-August-Universit\"at G\"ottingen, \\ Friedrich-Hund-Platz 1, D-37077 G\"ottingen, Germany\\
       E-mail: \email{mohammad.kareem@cern.ch}
       }

\abstract{Precise measurements of the properties of the top quark test the Standard Model (SM) and can be used to constrain new physics models. The top quark pair charge asymmetry is an asymmetry predicted to occur beyond leading-order QCD in the SM, and may be significantly enhanced by the presence of new physics.
The \ttbar production charge asymmetry is measured inclusively and differentially using the 8 TeV ATLAS dataset in the lepton+jets final state, including a dedicated measurement for highly boosted top quarks. The results are in agreement with the SM and compared to new physics models.
The top quark is predicted in the SM to decay almost exclusively to a \textit{W} boson and \textit{b} quark. Searches for non-SM top quark decays using the 8 TeV ATLAS dataset, including $t \rightarrow qH$ and $t \rightarrow qZ$ are presented. In addition, the measurement of the spin correlations in top quark pair production using the polar angles distribution is discussed.}

\FullConference{XXIV International Workshop on Deep-Inelastic Scattering and Related Subjects\\
                 11-15 April, 2016\\
                 DESY Hamburg, Germany}
\newcommand{\ttbar}{\ensuremath{t\bar{t}\ }}

\begin{document}

\section{Introduction}
Discovered in 1995 at the Tevatron $p\bar{p}$ collider by the CDF and D0 collaborations \cite{CDF,D0}, the top quark is known as the heaviest particle in the Standard Model (SM) of particle physics. In the Large Hadron Collider (LHC) \cite{LHC}, top quarks are produced in pairs through the strong interaction and individually through electroweak processes via proton-proton collisions. With a mass around 173 GeV close to the electroweak symmetry breaking scale, measurements of the top quark properties can provide an important tool in terms of tests of the Standard Model. The top quark has an extremely short life time that leads to a decay before hadronization, providing a unique opportunity to study the bare quark properties. Due to the high production rate of the top quark at the LHC, top quark properties measurements are of great importance.

In this context, using data taken at centre-of-mass energies of $\sqrt{s}$ = 7 TeV and 8 TeV, a variety of analyses are carried out by the ATLAS collaboration \cite{ATLAS} where some of the most recently obtained results are discussed here.



\section{Charge asymmetry}
The production rate difference in positive and negative absolute rapidity difference between top quarks and top antiquarks known as charge asymmetry ($A_{C}$), defined as

\begin{equation}
A_{C} = \frac{N(\Delta |y| > 0) - N(\Delta |y| < 0)}{N(\Delta |y| > 0) + N(\Delta |y| < 0)} ,
\label{eq:AC}
\end{equation}
is one of the interesting features of \ttbar production in $p\bar{p}$ collisions. The Standard Model at next-to-leading order (NLO) in Quantum Chromodynamics (QCD) predicts a charge asymmetry at the level of $A_{C} \sim 1\%$. On the other hand, several processes beyond the Standard Model (BSM) can alter $A_{C}$, either with anomalous vector or axial-vector couplings (e.g. axi-gluons) or via interference with SM processes, predicting different asymmetries.

One of the latest charge asymmetry measurements at ATLAS uses the full $\sqrt{s}$ = 8 TeV data set corresponding to an integrated luminosity of 20.3 fb$^{-1}$ in the single-lepton final state (where one top quark decays to a leptonically decaying \textit{W} boson and the other top quark decays to a hadronically decaying \textit{W} boson). The result was obtained inclusively and differentially as a function of invariant mass, transverse momentum and longitudinal boost ($\beta_{Z}$) of the \ttbar system, requiring at least four jets, one high $p_{T}$ lepton and missing transverse energy. The events are reconstructed via a kinematic likelihood fit method~\cite{klfitter} and a Bayesian unfolding procedure is applied
to account for distortions due to the acceptance and detector effects, leading to parton-level $A_{C}$ measurements. The inclusive measurement yields a value of $A_{C} =0.009 \pm 0.005$ (stat. + syst.)~\cite{CA}.

In addition, ATLAS performed the charge asymmetry measurement in the boosted topology \cite{CA_boost}, by reconstructing the hadronic hemisphere of the \ttbar event as one large-radius jet with radius parameter $R=1.0$ and $p_{T} > 300$ GeV inclusively and differentially. The boosted topology provides an accurate $A_{C}$ measurement as a function of the \ttbar invariant mass ($m_{\ttbar}$) in the TeV range by a more precise reconstruction of $m_{\ttbar}$ for events with highly boosted top quarks. The inclusive measurement for $m_{\ttbar} > $0.75 TeV and $|\Delta |y|| < 2$ yields $A_{C}=(4.2 \pm 3.2)\% $, that is within one standard deviation of the SM expectation. Furthermore, the differential measurement as function of the invariant mass of the \ttbar system disfavours t-channel W$^{'}$ boson model\cite{AguilarSaavedra:2011hz} in the highest $m_{\ttbar}$ bin.

\begin{figure}[h]
\begin{center}
		\includegraphics[height=55mm]{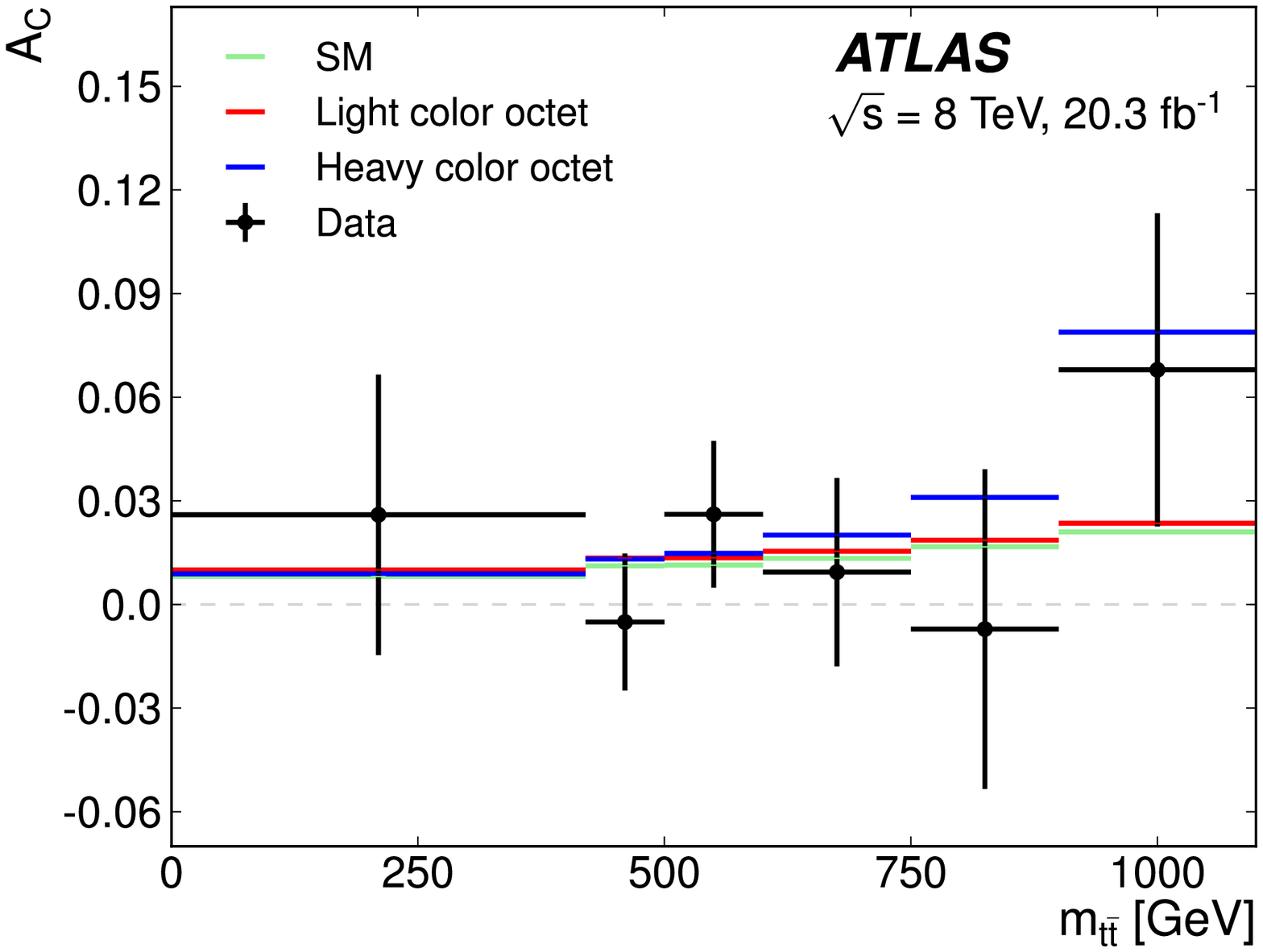}
		\includegraphics[height=56mm]{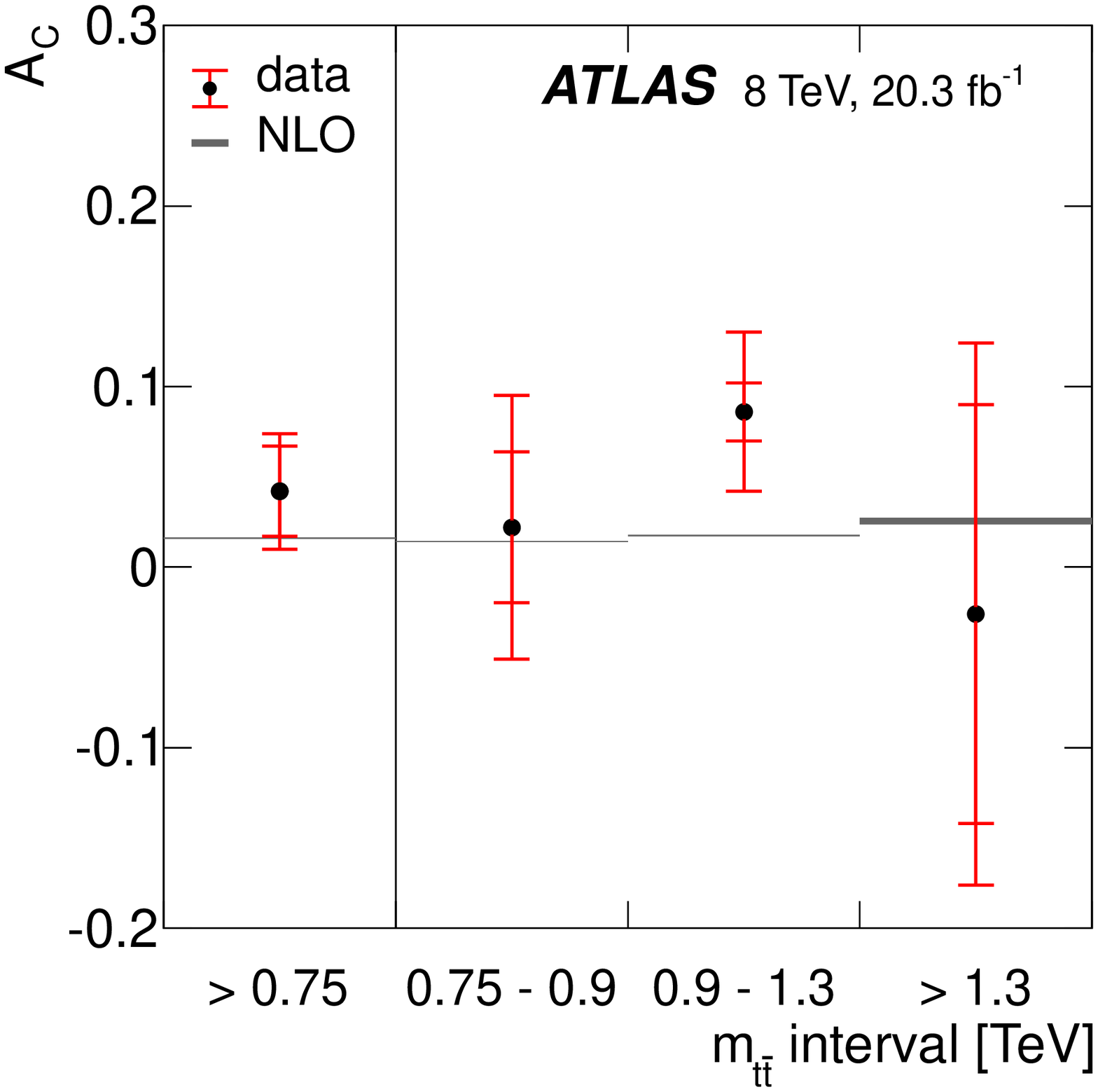}
	\caption{The $A_{C}$ measurement in resolved topology\cite{CA} (left), compared with predictions for SM and for right-handed colour octets with masses below the \ttbar threshold and beyond the kinematic reach of current LHC searches and boosted topology\cite{CA_boost} (right) as a function of $m_{\ttbar}$ compared with the SM prediction of the NLO calculation.}
	\label{fig:AC}
\end{center}	
\end{figure}

Figure \ref{fig:AC} shows the differential $A_{C}$ measurement in resolved and boosted topologies as a function of $m_{\ttbar}$. These measurements provide a constraint on extensions of the SM. The results of both measurements agree with the SM prediction.



\section{Rare decays of the top quark}
Within the Standard Model, flavour changing neutral currents (FCNC) are forbidden at tree level and are heavily suppressed via the GIM mechanism \cite{GIM}. In contrast, many BSM models predict significant enhancements at the level of the experimental accessibility, making this top quark property measurements an area of high interest. Within this context, ATLAS performed various searches for FCNC processes such as recent searches for $\mathscr{B}(t \rightarrow qH)$ \cite{FCNC_qH} and $\mathscr{B}(t \rightarrow qZ)$ \cite{FCNC_qZ}, using \ttbar events produced in the full $\sqrt{s}$ = 8 TeV data set corresponding to an integrated luminosity of 20.3 fb$^{-1}$, with one top quark decaying through the FCNC mode and the other through the SM dominant mode ($t \rightarrow bW$). Only the decays of the Higgs boson to $b\bar{b}$ and the Z boson to charged leptons and leptonic \textit{W} boson decays are considered, respectively. The final state topology of the top decay through the $tqH$ process  is characterised by an isolated high transverse momentum lepton and at least four jets. The final state characteristics of top quark decays through the $tqZ$ process is given by three isolated charged leptons, at least two jets, and missing transverse momentum from the undetected neutrino.

\begin{figure}[ht]
\begin{center}
		\includegraphics[height=55mm]{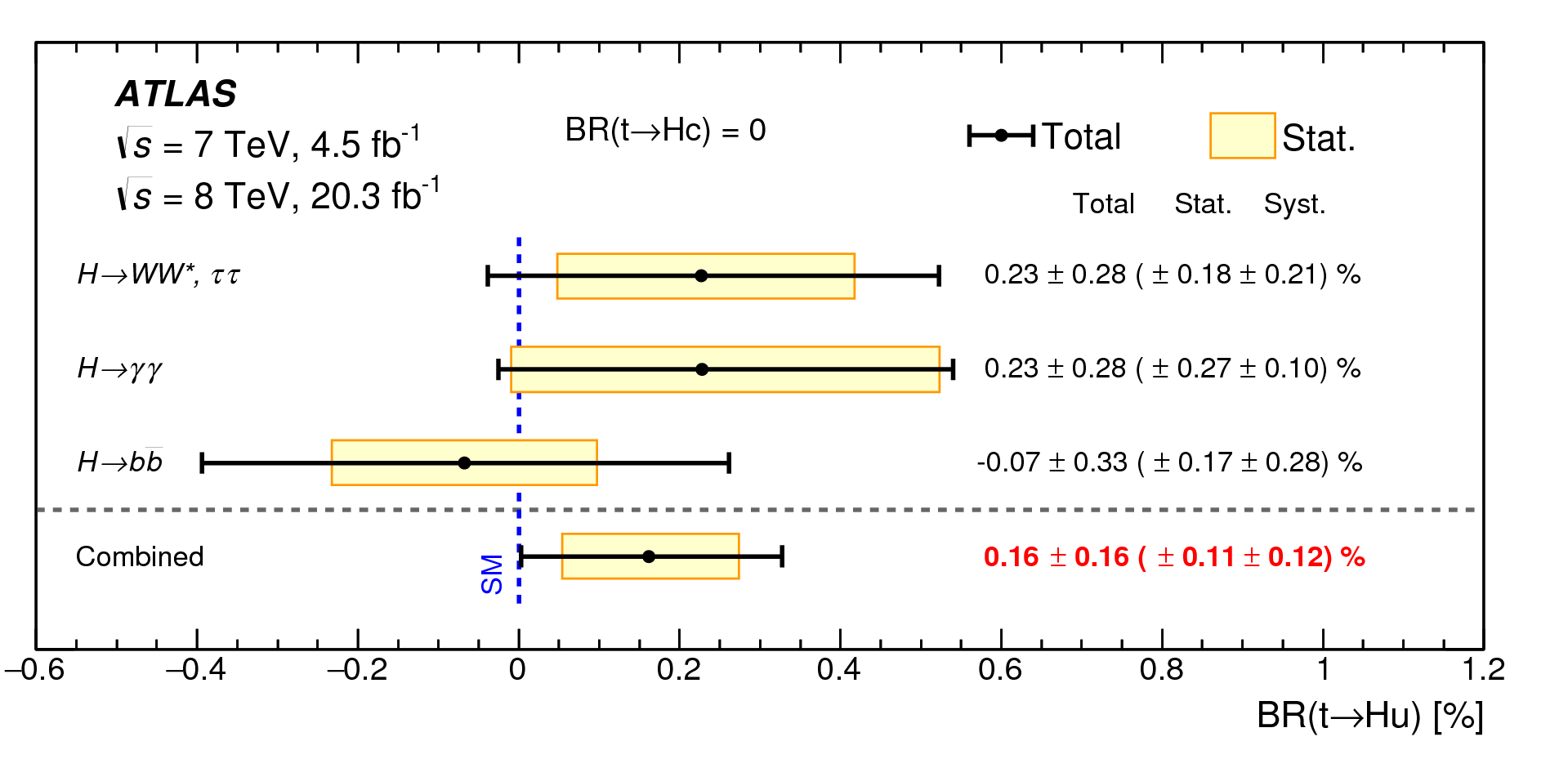}
		\includegraphics[height=55mm]{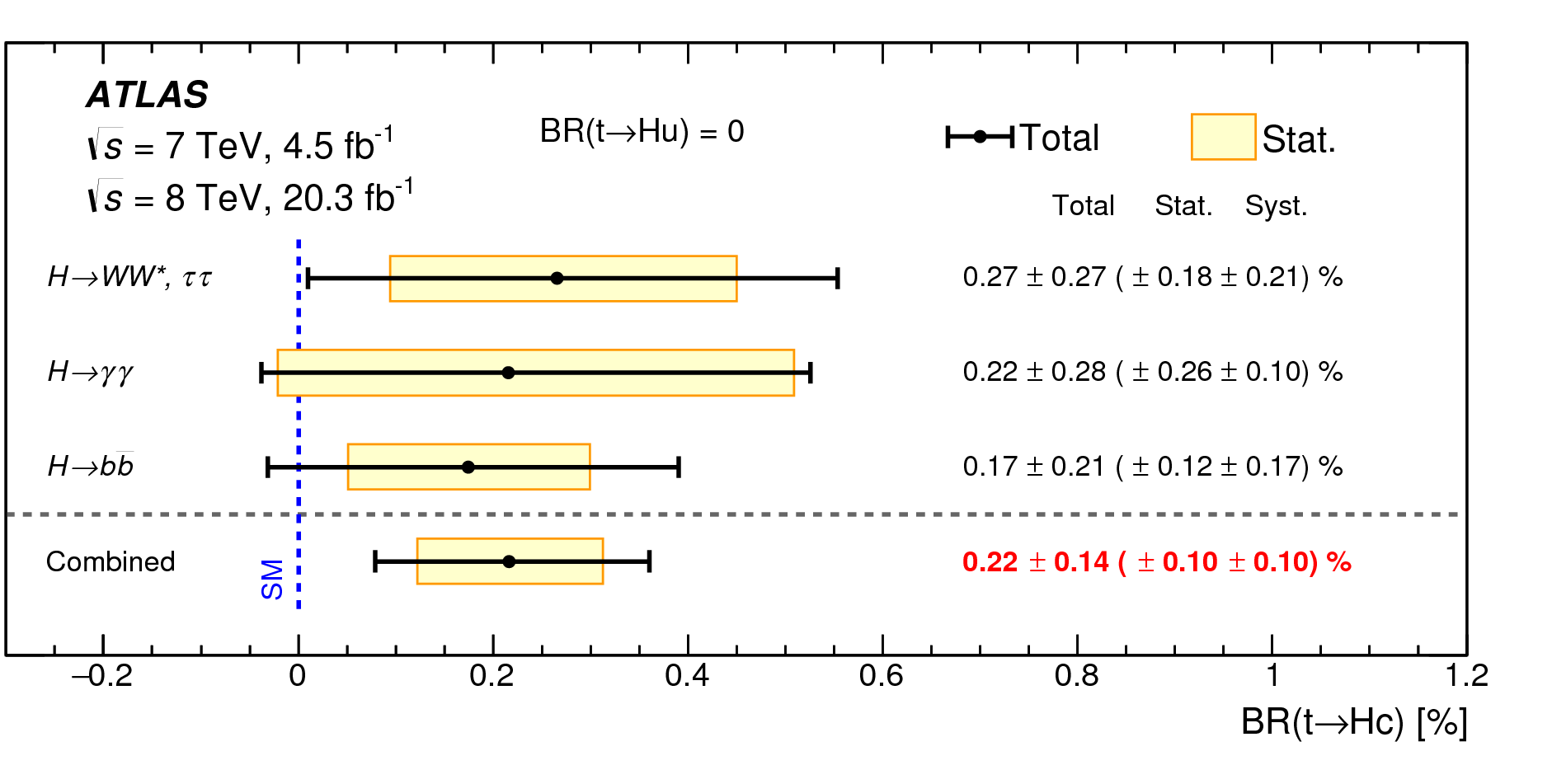}
	\caption{The best-fit for the individual searches as well as their combination for (top) $\mathscr{B}(t \rightarrow Hu)$, assuming that $\mathscr{B}(t \rightarrow Hc) =0$ and (bottom) $\mathscr{B}(t \rightarrow Hc)$, assuming that $\mathscr{B}(t \rightarrow Hu) =0$ \cite{FCNC_qH}.}
	\label{fig:FCNC_comined}
\end{center}	
\end{figure}

For the $tqH$ process, results from other ATLAS searches with $H \rightarrow \gamma \gamma$ and $H \rightarrow W^{+}W^{-}, \tau^{+}\tau^{-}$ have been combined with $H \rightarrow b\bar{b}$ to obtain the most restrictive direct bounds on $tqH$ interactions measured so far. Figure \ref{fig:FCNC_comined} summarises the best fit for the individual searches as well as their combination.

No evidence for signal events above the background expectation is found, but the established limits for these FCNC processes are in agreement with the expected theoretical limits.

\section{Spin correlation}
Because of the extreme short lifetime of the top quark, it decays before hadronization and that implies that the spin information of the top quark can be accessed from the angular momentum distributions of its decay products. The degree of correlation between the spin of the top quark and the top anti-quark is sensitive to the production mechanism. However, many scenarios of physics beyond the Standard Model predict different spin correlations, e.g. models including axi-gluons, $W^{'}$ bosons, extra right handed top-quark coupling, etc.

\begin{figure}[ht]
\begin{center}
		\includegraphics[height=85mm]{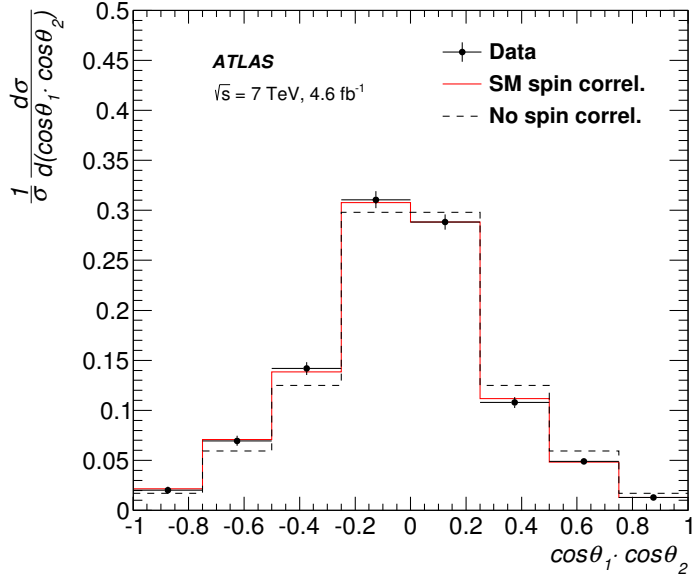}
	\caption{The unfolded data distribution of $\cos \theta_{1}.\cos \theta_{2}$. The predictions from the SM and the MC@NLO sample without spin correlation are overlaid for comparison. A symmetric distribution around zero would indicate no spin correlation \cite{Spin}.} 
	\label{fig:Spin}
\end{center}	
\end{figure}

A recent measurement of the correlations between the polar angles of leptons from the decay of top quarks in \ttbar events in the helicity basis carried out at the ATLAS experiment \cite{Spin}. The data set corresponds to an integrated luminosity of 4.6 fb$^{-1}$ at a centre-of-mass energy of $\sqrt{s}$ = 7 TeV with candidate events selected in the dilepton topology (where both top quarks in the \ttbar event decay to a leptonically decaying \textit{W} boson) with large missing transverse momentum and at least two jets. The angles $\theta_{1}$ and $\theta_{2}$ between the charged leptons and the direction of motion of the parent quarks in the \ttbar system rest frame are sensitive to the spin information, and the distribution of $\cos \theta_{1}.\cos \theta_{2}$ is sensitive to the spin correlation between the top quark and the top anti-quark. The events are reconstructed via the so called topology reconstruction method and the result is unfolded to the parton level using a fully Bayesian unfolding algorithm. The unfolded distribution is in good agreement with the prediction from MC@NLO \cite{MC_NLO} as displayed in Fig. \ref{fig:Spin}. 

In terms of $A_{helicity}=(N_{like}-N_{unlike})/(N_{like}+N_{unlike})$, where $N_{like}$ ($N_{unlike}$) is the number of events where the spins of the top quark and top anti-quark are (anti-)parallel with respect to the helicity basis, the result yields a value of $A_{helicity}= 0.315 \pm 0.061(stat.) \pm 0.049(syst.)$, and is in a good agreement with the NLO QCD prediction of   $A_{helicity}= 0.31$ \cite{NLO_correlation}.

\section{Conclusion}
Measuring top quark properties with high precision is a very good probe for the Standard Model and could potentially open a window to beyond Standard Model physics. 
A variety of recent measurements of top quark properties at ATLAS are discussed. Inclusive and differential charge asymmetry measurements in resolved and boosted topologies in the single lepton final state at a centre-of-mass energy of $\sqrt{s}$ = 8 TeV are presented. The measurement with boosted topology extended the reach of previous ATLAS and CMS\cite{CMS_CA} measurements to beyond 1 TeV and disfavours a t-channel W$^{'}$ boson model in the highest $m_{\ttbar}$ bin. However, no significant deviation from the SM expectation is observed. Having the data statistics as the limiting factor in these measurements, analysing the LHC Run II data with higher centre-of-mass energy of collision and luminosity will be of great interest. 
Furthermore, several searches for the observation of the flavour changing neutral current processes carried out with no significant evidence. An increased amount of data and improved measurement techniques will soon improve the current limits. The measurement of the correlations between the polar angles of leptons from the decay of the pair-produced top quark and top antiquark in the helicity basis in dilepton final state at a centre-of-mass energy of $\sqrt{s}$ = 7 TeV confirms the spin correlation in top quark decay and is in good agreement with the predictions of the Standard Model.



\bibliographystyle{JHEP}%
\bibliography{kareem}%

\providecommand{\href}[2]{#2}\begingroup\raggedright\begin{thebibliography}{10}

\bibitem{CDF}
{\scshape CDF} collaboration, F.~Abe et~al., \emph{{Observation of top quark
  production in $\bar{p}p$ collisions}}, {\emph{Phys. Rev. Lett.} {\bf 74}
  (1995) 2626}.

\bibitem{D0}
{\scshape D0} collaboration, S.~Abachi et~al., \emph{{Observation of top
  quark}}, {\emph{Phys. Rev. Lett.} {\bf 74} (1995) 2632}.

\bibitem{LHC}
L.~Evans and P.~Bryant, \emph{{LHC Machine}},
  \href{http://dx.doi.org/10.1088/1748-0221/3/08/S08001}{\emph{JINST} {\bf 3}
  (2008) S08001}.

\bibitem{ATLAS}
{ATLAS Collaboration}, \emph{{The ATLAS Experiment at the CERN Large Hadron
  Collider}},
  \href{http://dx.doi.org/10.1088/1748-0221/3/08/S08003}{\emph{JINST} {\bf 3}
  (2008) S08003}.

\bibitem{klfitter}
J.~Erdmann et~al., \emph{{A likelihood-based reconstruction algorithm for
  top-quark pairs and the KLFitter framework}},
  \href{http://dx.doi.org/10.1016/j.nima.2014.02.029}{\emph{Nucl. Instrum.
  Meth.} {\bf A748} (2014) 18--25}.

\bibitem{CA}
{\scshape ATLAS} collaboration, \emph{{Measurement of the charge asymmetry in
  top-quark pair production in the lepton-plus-jets final state in pp collision
  data at $\sqrt{s}=8\,\mathrm TeV{}$ with the ATLAS detector}},
  \href{http://dx.doi.org/10.1140/epjc/s10052-016-3910-6}{\emph{Eur. Phys. J.}
  {\bf C76} (2016) 87}.

\bibitem{CA_boost}
{\scshape ATLAS} collaboration, \emph{{Measurement of the charge asymmetry in
  highly boosted top-quark pair production in $\sqrt{s} =$ 8 TeV $pp$ collision
  data collected by the ATLAS experiment}},
  \href{http://dx.doi.org/10.1016/j.physletb.2016.02.055}{\emph{Phys. Lett.}
  {\bf B756} (2016) 52--71}.

\bibitem{AguilarSaavedra:2011hz}
J.~A. Aguilar-Saavedra and M.~Perez-Victoria, \emph{{Asymmetries in t $\bar{t}$
  production: LHC versus Tevatron}},
  \href{http://dx.doi.org/10.1103/PhysRevD.84.115013}{\emph{Phys. Rev.} {\bf
  D84} (2011) 115013}.

\bibitem{GIM}
S.~L. Glashow, J.~Iliopoulos and L.~Maiani, \emph{Weak interactions with
  lepton-hadron symmetry},
  \href{http://dx.doi.org/10.1103/PhysRevD.2.1285}{\emph{Phys. Rev. D} {\bf 2}
  (1970) 1285}.

\bibitem{FCNC_qH}
{\scshape ATLAS} collaboration, \emph{{Search for flavour-changing neutral
  current top quark decays $t\to Hq$ in $pp$ collisions at $\sqrt{s}=8$ TeV
  with the ATLAS detector}},
  \href{http://dx.doi.org/10.1007/JHEP12(2015)061}{\emph{JHEP} {\bf 12} (2015)
  061}.

\bibitem{FCNC_qZ}
{\scshape ATLAS} collaboration, \emph{{Search for flavour-changing neutral
  current top-quark decays to $qZ$ in $pp$ collision data collected with the
  ATLAS detector at $\sqrt s =8$ TeV}},
  \href{http://dx.doi.org/10.1140/epjc/s10052-015-3851-5}{\emph{Eur. Phys. J.}
  {\bf C76} (2016) 12}.

\bibitem{Spin}
{\scshape ATLAS} collaboration, \emph{{Measurement of the correlations between
  the polar angles of leptons from top quark decays in the helicity basis at
  $\sqrt{s}=7$TeV using the ATLAS detector}},
  \href{http://dx.doi.org/10.1103/PhysRevD.93.012002}{\emph{Phys. Rev.} {\bf
  D93} (2016) 012002}.

\bibitem{MC_NLO}
S.~Frixione, P.~Nason and B.~R. Webber, \emph{{Matching NLO QCD and parton
  showers in heavy flavor production}},
  \href{http://dx.doi.org/10.1088/1126-6708/2003/08/007}{\emph{JHEP} {\bf 08}
  (2003) 007}.

\bibitem{NLO_correlation}
W.~Bernreuther, A.~Brandenburg, Z.~G. Si and P.~Uwer, \emph{{Top quark spin
  correlations at hadron colliders: Predictions at next-to-leading order QCD}},
  \href{http://dx.doi.org/10.1103/PhysRevLett.87.242002}{\emph{Phys. Rev.
  Lett.} {\bf 87} (2001) 242002}.

\bibitem{CMS_CA}
{\scshape CMS} collaboration, V.~Khachatryan et~al., \emph{{Measurement of the
  charge asymmetry in top quark pair production in pp collisions at $\sqrt(s)
  =$ 8 TeV using a template method}},
  \href{http://dx.doi.org/10.1103/PhysRevD.93.034014}{\emph{Phys. Rev.} {\bf
  D93} (2016) 034014}.

\end{thebibliography}\endgroup

\end{document}